\documentclass[prl, nofootinbib, floatfix, notitlepage, twocolumn]{revtex4-1}

\usepackage{amsmath,amsfonts,amssymb}
\usepackage{mathrsfs}
\usepackage{graphicx}
\usepackage[english]{babel} 
\usepackage{hyperref} 
\usepackage{color}

\hypersetup{
    colorlinks=true,
    linkcolor=red,
    citecolor=blue,
} 

\newcommand{\be}{\begin{equation}}
\newcommand{\ee}{\end{equation}}
\newcommand{\bes}{\begin{equation*}}
\newcommand{\ees}{\end{equation*}}
\newcommand{\mn}{{\mu \nu}}

\begin{document}

\title{Chiral Imprint of a Cosmic Gauge Field on Primordial Gravitational Waves}
\author{Jannis Bielefeld}
\author{Robert R. Caldwell}
\affiliation{Department of Physics and Astronomy, Dartmouth College, 6127 Wilder Laboratory, Hanover, New Hampshire 03755 USA}

\date{\today}


\begin{abstract}
A cosmological gauge field with isotropic stress-energy introduces parity violation into the behavior of gravitational waves. We show that a primordial spectrum of inflationary gravitational waves develops a preferred handedness, left- or right-circularly polarized, depending on the abundance and coupling of the gauge field during the radiation era.  A modest abundance of the gauge field would induce parity-violating correlations of the cosmic microwave background temperature and polarization patterns that could be detected by current and future experiments.
\end{abstract}
\maketitle

Since the surprising discovery that parity is violated on the atomic scale by the weak nuclear force \cite{Wu:1957my}, searches for broken symmetries have proven to be a remarkably effective technique for uncovering new laws of physics. Here we speculate that parity violation on cosmic scales may be the sign of a dark gauge field. Our theoretical model consists of a non-Abelian (Yang-Mills) gauge field which, as we demonstrate, behaves like radiation through cosmic history, but fluctuations of the field couple to gravity in a way that distinguishes between left- and right-handed circularly polarized gravitational waves. In this paper we demonstrate the effect of this field on a primordial spectrum of gravitational waves and evaluate its impact on the cosmic microwave background (CMB).  

\begin{figure}[t]
\includegraphics[width=1\linewidth]{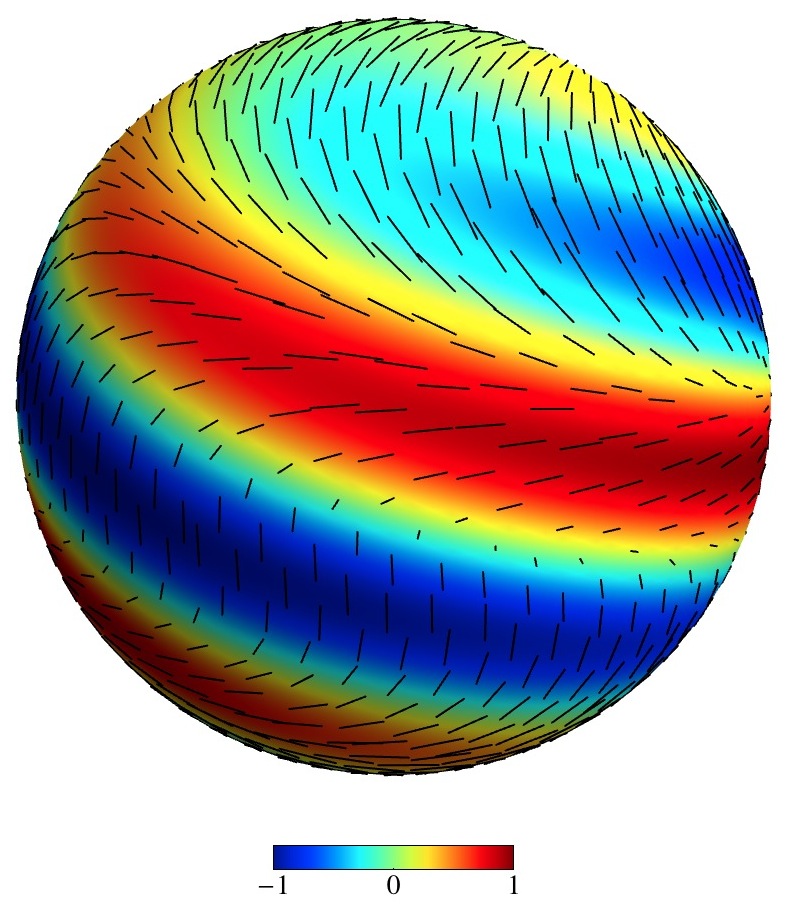}
\caption{A single right-handed circularly polarized gravitational wave, in this case with wavelength equal to the radius of the surface of last scattering and propagating in the vertical direction, leaves a left-handed spiral pattern of temperature anisotropy and polarization excess on the cosmic microwave background (CMB) sky. In our scenario, the parity reversed wave is damped by the flavor-space locked gauge field, so that the ``mirror image" is a right-handed spiral pattern of much weaker temperature and polarization. There is a curl pattern around the zero temperature swath passing diagonally across the equator.}
\label{fig:fig0}
\vspace{-0.75cm}
\end{figure}

There is an extensive literature on speculative new physics that leads to parity violation in cosmology, e.g. Refs.~\cite{Carroll:1989vb,Lue:1998mq,Contaldi:2008yz}, and more specifically in the gravitational sector \cite{Alexander:2009tp}. But an asymmetry is perfectly compatible with general relativity, without the need to invoke exotic interactions, as Stueckelberg first pointed out \cite{Stueckelberg:1957}. The physics of the weak interaction contains all the necessary elements. 

We consider the standard cosmological model with the sole addition of a new gauge field as a toy model, with an action given by
\be
\begin{gathered}
S = \int d^4 x \sqrt{-g} \left(\frac{1}{2} M_P^2 R + {\cal L}_m - \frac{1}{4} F_{I\mu\nu}F^{I\mu\nu}\right) \cr
F^{I}_\mn  \equiv  \partial_\mu A^{I}_\nu - \partial_\nu A^{I}_\mu -g_{\rm YM}\epsilon^{IJK}A_{J\mu} A_{K\nu}
\end{gathered}
\ee
where Greek letters are used to represent space-time indices, lower case Latin letters are spatial indices, and upper case Latin letters $I \in \{1,2,3\}$ are reserved for the SU(2) indices. We assume a {\it flavor-space locked} configuration for the gauge field, wherein $A_i^I = \phi(\tau) \delta^I_i$, so that the directions of the internal group space are aligned with the principle spatial axes of the Robertson-Walker space-time, $ds^2 = a^2(\tau)(-d\tau^2 + d\vec x^2)$. This approach yields an isotropic stress-energy tensor for the gauge field \cite{ArmendarizPicon:2004pm}. The scalar potential equation of motion is $\phi'' + 2 g_{\rm YM}^2 \phi^3=0$ where the prime indicates derivative with respect to conformal time, with a well-known solution in terms of the Jacobi elliptic sine-amplitude function \cite{ByrdFriedman}, $g_{\rm YM} \phi(\tau) = c_1 {\rm sn}(c_1(\tau-\tau_i) + c_2 | -1)$. The constants are determined at the initial time $\tau_i$ as $c_1^4 = g_{\rm YM}^2(\phi_i'^2 + g_{\rm YM}^2 \phi_i^4)$ and $c_2 = F(\theta|-1)$, an elliptic integral of the first kind, with $\csc\theta_i = 1 + \phi_i'^2/g_{\rm YM}^2 \phi_i^4$. The homogeneous, isotropic energy density and pressure of this Yang-Mills fluid are $\rho_{\rm YM} = 3 p_{\rm YM} =3(\phi'^2 + g_{\rm YM}^2 \phi^4)/2a^4 = 3 c_1^4/2 g_{\rm YM}^2 a^4$. The gauge field oscillates with period $\tau =\Gamma(\tfrac{1}{4})^2/\sqrt{2 \pi} c_1$, but its energy density and pressure scale with equation of state $w=1/3$, like radiation. In order that the field oscillate on a time scale to affect cosmological physics, the coupling must be exponentially small, $g_{\rm YM}\sim {\cal O}(H_0/M_P)\sim 10^{-60}$. An origin for this small coupling as well as the flavor-space locked configuration could be ascribed to an inflationary epoch, although that is beyond the scope of our present investigation. Inflationary scenarios based on a similar gauge field, but requiring higher order couplings to matter fields, have been studied elsewhere \cite{Maleknejad:2011jw,Adshead:2012kp}. 
 
The parity violation is manifest in the coupling to gravitational waves. To explain, consider a gravitational wave passing through the gauge field described above. The wave will induce a quadrupolar distortion, alternately squeezing and stretching the stress and energy of the gauge field. However, the gauge field itself possesses a preferred handedness via the right-handed SU(2) structure constants. Since the fields are flavor-space locked, the gauge field stress-energy will vibrate in sympathy to a right-handed wave, and with antipathy to a left-handed wave, somewhat like a rattleback top \cite{rattleback}. In detail, we perturb the metric and gauge field $\delta g_{\mu\nu} = a^2(\tau) h_{\mu\nu}$, $\delta A_{I\mu} = a(\tau) M_P t_{ij}\delta^i_I \delta^j_\mu$  where $t_{ij}$ and $h_{ij}$ are transverse, traceless, synchronous tensors, following Refs.~\cite{Dimastrogiovanni:2012ew,Namba:2013kia}. The equations of motion for the Fourier amplitudes of a right circularly polarized gravitational wave traveling in the $+z$ direction, and the corresponding gauge field fluctuation, are given by
\begin{eqnarray}
&&h_R''+ 2 \frac{a'}{a}h_R' +\left[k^2  + \frac{2}{a^2 M_P^2}\left(g_{\rm YM}^2 \phi^4 - \phi'^2\right)\right] h_R =  \cr
&& \quad	-\frac{2}{a M_P}\left[ (k - g_{\rm YM} \phi) g_{\rm YM} \phi^2 t_R + \frac{a'}{a}\phi' t_R + \phi' t'_R\right], \nonumber \\
&&t_R''+2 \frac{a'}{a}t_R' + \left[k^2 + \frac{a''}{a} - 2 k g_{\rm YM} \phi \right] t_R  =  \cr
&& \quad	- \frac{2}{a M_P}\left[(k+g_{\rm YM} \phi) g_{\rm YM} \phi^2 h_R - \phi' h_R'\right]. 
	\label{eqn:Htens}
\end{eqnarray}
Upon the exchange $k \to -k$, the equations now describe a left circularly polarized gravitational wave, but with very different consequences for the evolution --- circular dichroism --- as can be seen by examining the change in the effective mass term $-2 g_{\rm YM} \phi$ for $t$ and the coupling between $h$ and $t$. For clarity, we have omitted the anisotropic shear contributed by other species, such as photons and neutrinos, although these effects are included in our cosmic microwave background (CMB) analysis. There is a rich variety of behavior in the evolution of this system, dependent upon the coupling $g_{\rm YM}$, the abundance $R_{\rm YM} = \rho_{\rm YM}/\rho_{\rm rad}$ during the radiation era, the relative contributions of electric and magnetic field energy, $\phi'$ and $g_{\rm YM} \phi^2$, and the initial conditions for the perturbations $h_{R/L},\, t_{R/L}$.

We investigate a minimal scenario in which the initial field energy of the YM fluid is split equally between the electric and magnetic field, $\phi' = g_{\rm YM} \phi^2$. We further assume equal amplitude scale-free primordial spectra of left- and right-handed gravitational waves. These initial conditions, and the assumption that the initial tensor fluctuations of the gauge field vanish deep in the radiation era, allows the YM fluid to behave like radiation at early times. The effects of the gauge field on the subsequent evolution of the gravitational waves are illustrated in the figures below.

\begin{figure}[t]
\includegraphics[width=1\linewidth]{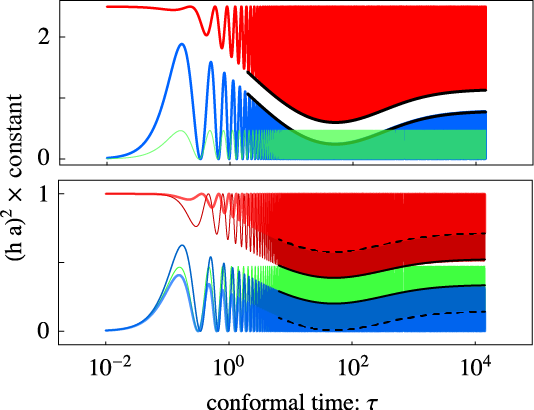}
\caption{Top: Gravitational wave amplitude evolution as a function of conformal time is shown for the case $g_{\rm YM}=0$, $R_{\rm YM}=0.03$ (blue) and wavenumber  $k=10$~h/Mpc, as compared to the standard case $R_{\rm YM}=0$ (green). The excitations of the gauge field are shown (red) as an offset minus a constant times $(a t_{A})^2$, to illustrate their complementary behavior. The solid (black) lines show the results of WKB solutions for the envelopes of the oscillatory waveforms. Bottom: The case $g_{\rm YM}=10^{-60}$, $R_{\rm YM}=0.03$ is shown. The waveforms capped with solid lines are right-handed; those with dashed lines are left-handed.}
\label{fig:fig1}
\vspace{-0.5cm}
\end{figure}

\begin{figure}[t]
\includegraphics[width=0.99\linewidth]{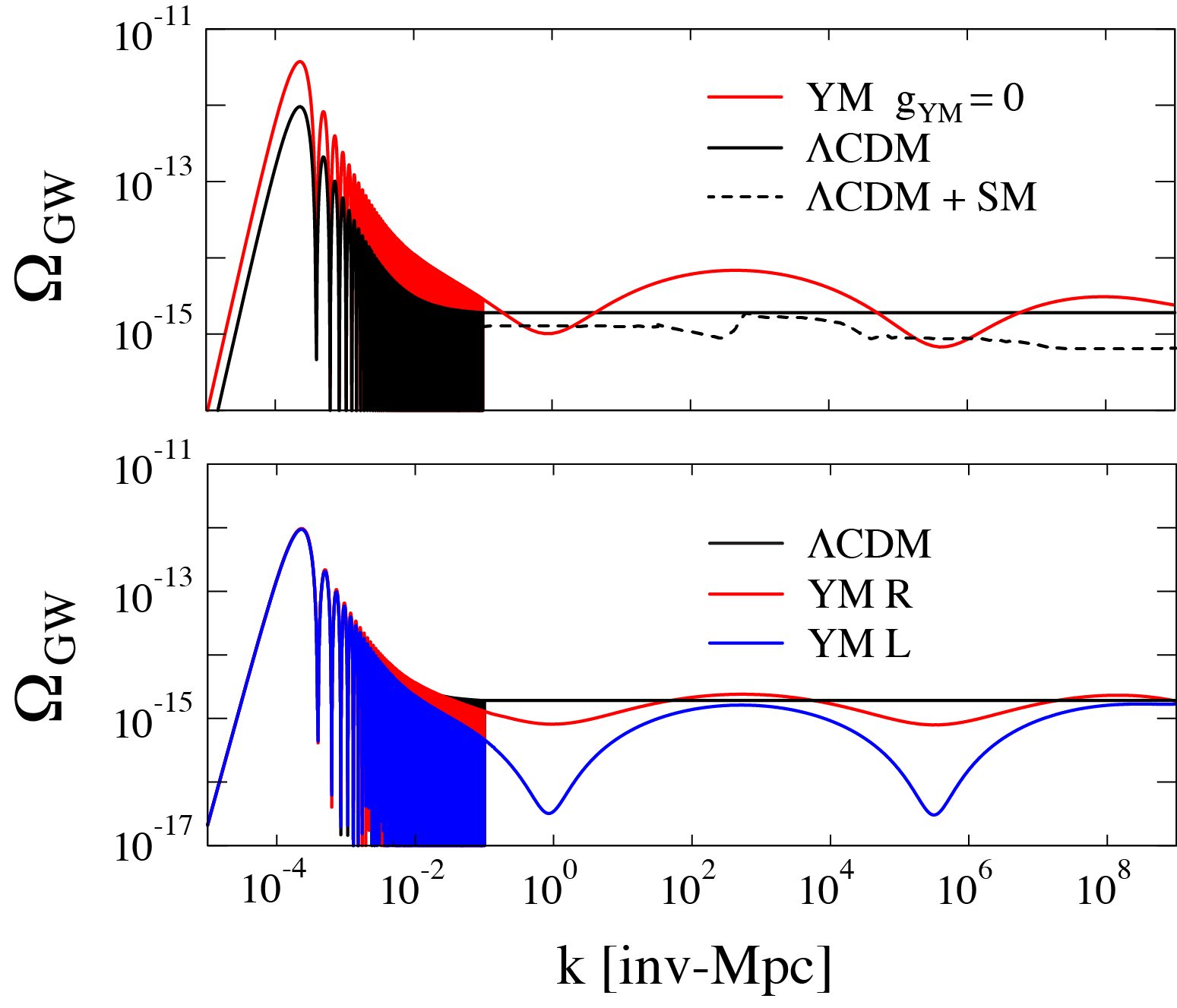}
\caption{The gravitational wave energy density spectrum is shown as a function of comoving wavenumber. An ambidextrous, scale free spectrum at an inflationary scale $H_I = 10^{-5}M_P$ is assumed. Top: The present-day spectrum in the case with $g_{\rm YM}=0$, $R_{\rm YM}=0.03$ displays large oscillatory features due to the coupling between the gravitational waves and the gauge field. For comparison, the standard case without the gauge field is shown, as well as the effect of Standard Model particle free-streaming and freeze-outs (taken from Ref.~\cite{Watanabe:2006qe}). Bottom: The case $g_{\rm YM}=10^{-60}$, $R_{\rm YM}=0.03$ is shown. The long wavelength modes are unaffected given our choice of initial conditions in the background field.}
\label{fig:fig2}
\vspace{-0.5cm}
\end{figure}

The evolution of the gravitational wave amplitude is shown for a variety of cases in Fig.~\ref{fig:fig1}. We begin by examining the behavior in the case $g_{\rm YM}=0$, corresponding to color electrodynamics. The background solution has $\phi'$ constant, so the $h_{A}$ evolution equation (where the subscript ``A" is for ambidextrous, since there is no parity violation in this case) has a tachyonic mass that is responsible for the growth of long wavelength modes. One can show analytically that $a h_A$ doubles for modes outside the horizon, relative to the standard case. In the top panel of Fig.~\ref{fig:fig1}, the amplitude of $(a h_{A})^2$ (blue) is $2^2$ times the amplitude of $(a h)^2$ (green) going in to the first oscillation. For modes that enter the horizon, there is a slow exchange of amplitude between $h_{A}$ and $t_{A}$. A WKB analysis for sub-horizon modes shows that $h_A^2 + t_A^2 \propto 1/a^2$ and the exchange is oscillatory with phase $a_0 k_A\int d\tau' /a(\tau')$ where $k_A = \sqrt{2 R_{\rm YM} \Omega_{\rm rad}} a_0 H_0$ \cite{BieCald}. The gravitational wave spectral density $\Omega_{\rm GW}$ is shown in Fig.~\ref{fig:fig2}, where the amplification and periodic modulation are clearly seen. The WKB solution predicts a peak or dip in the spectral density every five orders of magnitude in $k$ for $R_{\rm YM}=0.03$ (an energy density comparable to $\Delta N_\nu \simeq 0.2$.). Modes that spend very little time outside the horizon after inflation experience very little growth, leading to a decay in the amplitude of modulation. Since we begin our numerical calculation at $a/a_0\sim 10^{-16}$, much later than expected in a typical inflationary scenario, this effect can be seen as the artificial decay of the spectral density at high frequency in the figure. Note that the influence of the gauge field on the spectrum is much larger than the effect of photon and neutrino free-streaming, or brief departures from a pure radiation background when particle species become non-relativistic.

The YM fluid case is shown in the second set of panels of Figs.~\ref{fig:fig1}, \ref{fig:fig2}. The effective (squared) mass term for the $h_{R/L}$ evolution equation now depends on the sign of $g_{\rm YM}^2\phi^4-\phi'^2$. If positive it will damp long wavelength modes, whereas if negative it will enhance long wavelength modes as in the $g_{\rm YM}=0$ case. Initial conditions for $\phi$ can be chosen to accommodate either sign without affecting the total energy density or pressure. However we fix the initial conditions so that the field has equal energy invested in the electric and magnetic modes, $g_{\rm YM}^2\phi^4=\phi'^2$, whereupon the effective mass vanishes at the starting point. The subsequent background evolution causes $(g_{\rm YM}^2\phi^4-\phi'^2)/a^4$ to become non-zero and grow relative to the energy density $\rho_{\rm YM}$ in a time scale given by the period of oscillation of $\phi$. If the coupling $g_{\rm YM}$ is sufficiently small, as for the case illustrated in Fig.~\ref{fig:fig2}, then this time scale is longer than the age of the universe, with the result that super horizon modes are unaffected by the gauge field. If the coupling is large enough so that the oscillations begin before the present day, then the effective mass will lead to a suppression of order unity, depending on the abundance $R_{\rm YM}$, of all super horizon modes. We note that such suppression, which can also be achieved through a different choice of initial conditions for $\phi,\,\phi'$ could modify the predictions of the gauge-flation and chromo-natural inflation scenarios \cite{Maleknejad:2011jw,Adshead:2012kp,Namba:2013kia}.

The difference in the evolution for left- and right-circularly polarized waves is primarily due to the effective mass term for the gauge field tensor perturbations, $-2 k g_{\rm YM}\phi$, which is tachyonic for right-handed modes. The growth (suppression) of $t_{R}$ ($t_{L}$) is transferred to $h_{R}$ ($h_{L}$) as the mode enters the horizon. Once the relative amplitude is locked in at horizon entry and the fields begin to oscillate rapidly, the slow exchange of amplitude between $h_{R/L}$ and $t_{R/L}$ again comes into play. A WKB analysis for sub-horizon modes again shows that $h_{R/L}^2 + t_{R/L}^2 \propto 1/a^2$ and the exchange is oscillatory with similar phase if the background field is not yet oscillatory. 
 
To evaluate the impact of this scenario on the CMB, we have implemented the scalar and tensor perturbations of the gauge field into CAMB \cite{Lewis:1999bs}. The gauge field has the biggest impact on tensor correlations. The scalar sector also receives corrections due to the gauge field, in the form of an anisotropic scalar shear, but the impact on the CMB scalar spectrum is small. We ignore the vector perturbations which may be shown to decay \cite{BieCald}.   For these calculations $R_{\rm YM}$ is the ratio between the YM fluid density and the total relativistic energy density. We assume the fraction of critical density in the relativistic fluid is fixed by slightly adjusting the number of neutrino degrees of freedom upon introducing the gauge field. We otherwise assume standard $\Lambda$CDM parameters.

\begin{figure}[tbp]
	\includegraphics[width=1\linewidth]{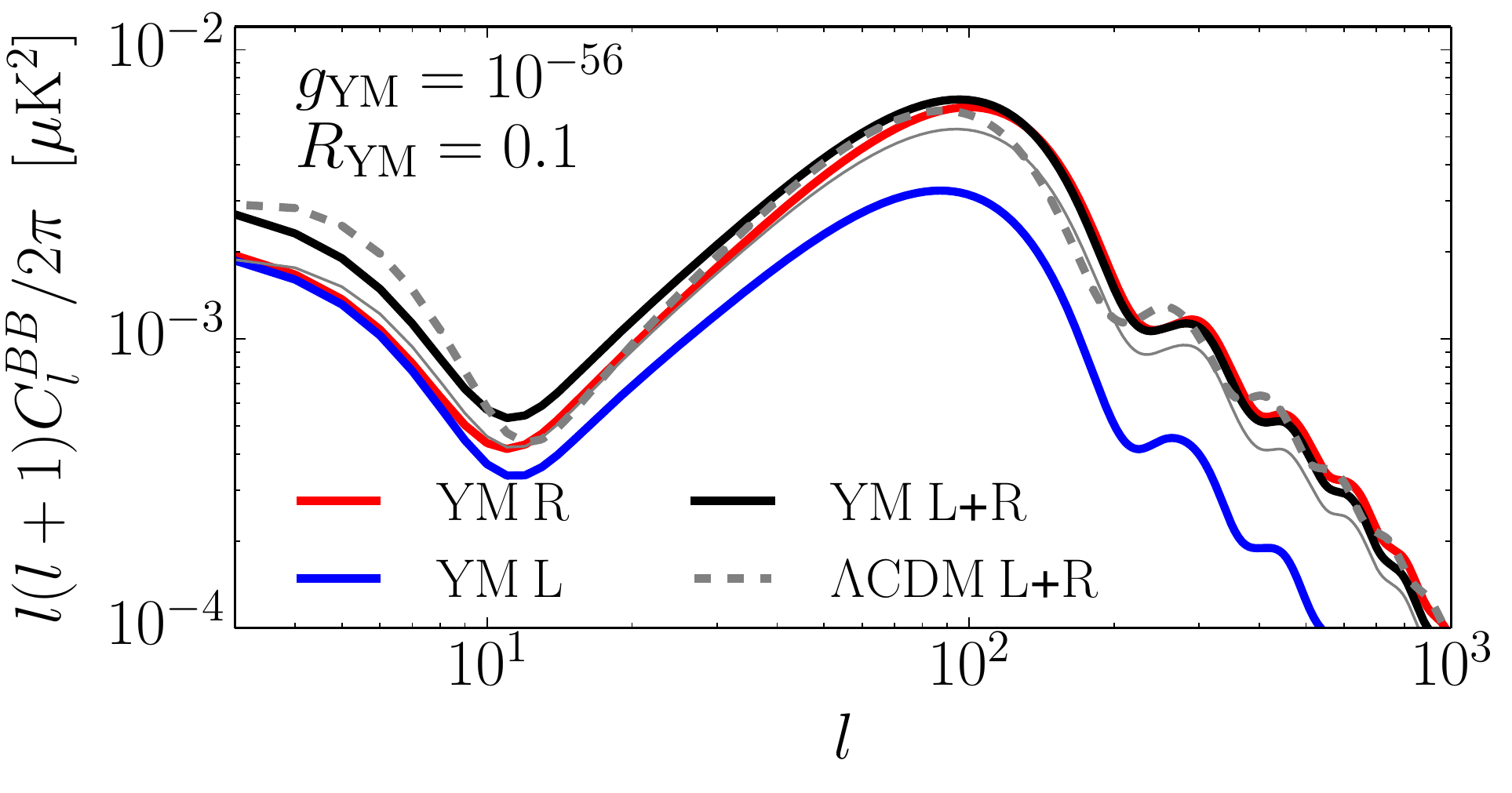}
	\caption{CMB $BB$ polarization autocorrelation spectra. The pure left- and right-handed contributions deviate strongly from $\Lambda$CDM (dashed) while their sum is closer to it. Solid lines include a gauge field with $g_{\rm YM} = 10^{-56}$, $R_{\rm YM} = 0.1$ and tensor-to-scalar ratio of $r = 0.1$. The thin line shows the $BB$ spectrum for the case $g_{\rm YM} = 4 \times10^{-56}$.}
	\label{fig:BB}
\vspace{-0.5cm}
\end{figure}

\begin{figure}[thbp]
	\includegraphics[width=1\linewidth]{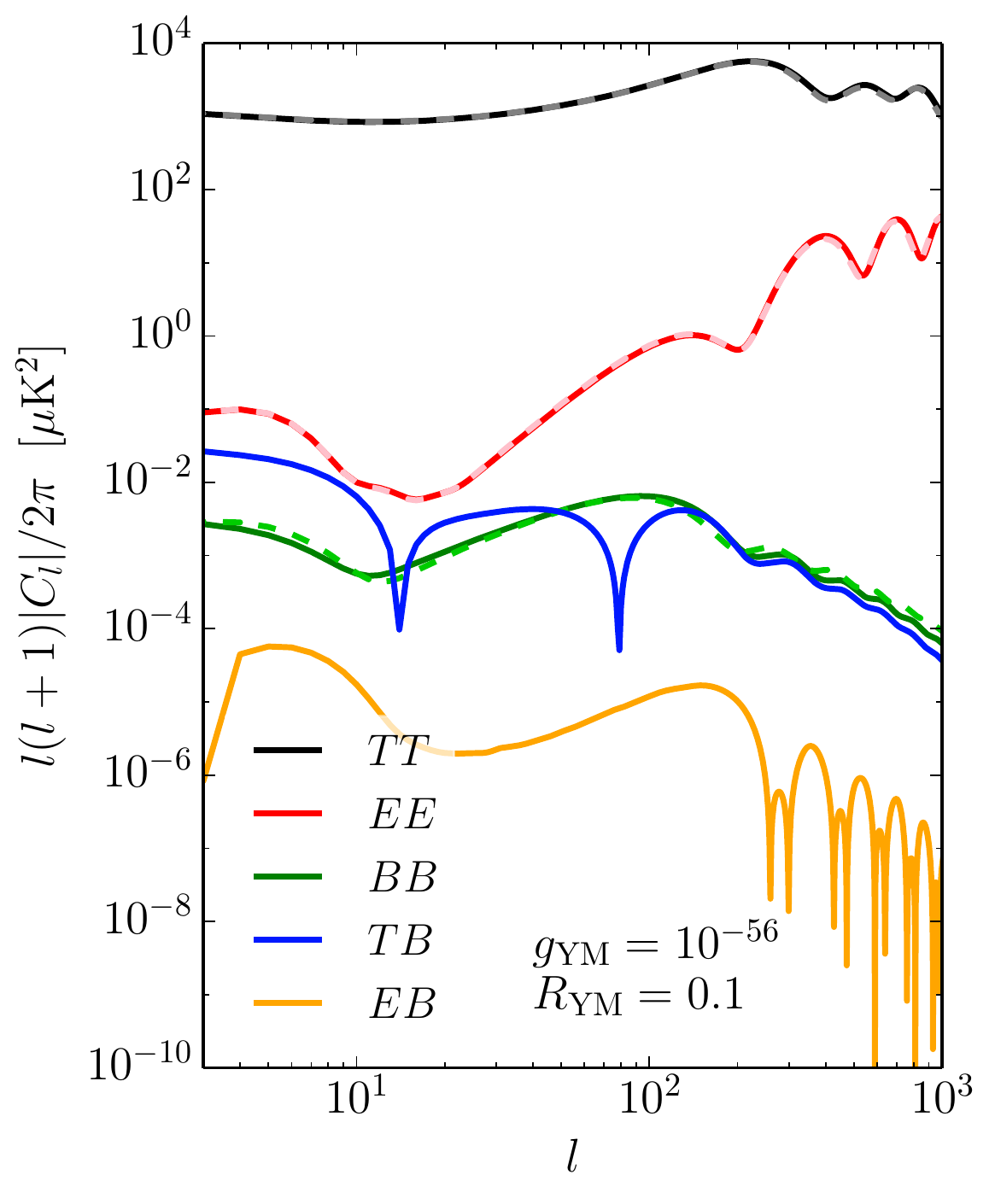}
	\caption{CMB temperature and polarization auto- and cross-correlation power spectra. The parity violation allows for $TB$ and $EB$ cross correlations. The dashed lines represent the standard $\Lambda$CDM cosmology, solid lines include a gauge field with $g_{\rm YM} = 10^{-56}$ and $R_{\rm YM} = 0.1$ and tensor-to-scalar ratio of $r = 0.1$. The $TB$ and $EB$ cross correlations only appear when the gauge field is present.}
	\label{fig:tens}
\vspace{-0.5cm}
\end{figure}

The CMB polarization can be decomposed into gradient $E$-modes and curl $B$-modes. In the tensor sector the gauge field introduces two main effects. First, the left- and right-handed contributions to the $BB$ spectrum now differ, as shown in Fig.~\ref{fig:BB}. Hence, the temperature and polarization anisotropy due to gravitational waves on roughly degree scales is dominated by a superposition of right-circularly polarized gravitational waves, which imprint left-helical patterns, similar to the display in Fig.~\ref{fig:fig0}. It is curious to see that the individual contributions deviate strongly from $\Lambda$CDM but conspire in a way that puts the combination of both close to the expected standard cosmology result. Second, because temperature $T$ and gradient polarization $E$ are both parity even but curl polarization is parity odd, the parity violation introduced by the YM fluid allows for correlations between $TB$ and $EB$ \cite{Lue:1998mq}. Detecting these exotic cross correlations is a smoking gun for chiral effects in the universe. Typical predictions of our model are plotted in Fig.~\ref{fig:tens}. 

There are many challenges to detecting the parity-violating cross correlations, not to mention the $B$-mode signal. Galactic foregrounds, magnetic fields, weak lensing, and other systematic effects can all produce a false positive; fortunately there is no fundamental barrier that would prevent a detection that can distinguish a primordial signal. (See Ref.~\cite{Kaufman:2014rpa} for a recent summary.) But there are other phenomena that could produce a parity-violating signal. First of all, cosmological birefringence (CB) can lead to $TB$ and $EB$ power spectra by rotating $E$ into $B$ through a novel coupling between electromagnetism and a cosmic pseudoscalar such as quintessence \cite{Carroll:1998zi}.  A second possibility, broadly characterized as chiral gravity, posits a modification of gravity whereby an asymmetry between left- and right-circularly polarized waves is imprinted on the primordial spectrum. The third possibility, as we have shown, is essentially cosmic circular dichroism, whereby the asymmetry develops with time from an initially symmetric primordial spectrum.

Would an actual detection directly point to chiral symmetry breaking on cosmological scales? In Ref.~\cite{Gluscevic:2010vv} it was shown that the $TB$ and $EB$ spectra can be used to distinguish CB effects from chiral physics. As CB rotates the $E$ into a $B$ contribution the measured $B$ spectrum would resemble the $E$ one which makes this separation into CB and chirality effects feasible. In turn, putting limits on the amplitude of these spectra will put constraints on chiral physics in general and our model in particular. In what follows we compute the constraints that current and future CMB experiments would put on the parameters of our model under the assumption that $TB$ and $EB$ cross-correlations are measured. 
  
The deviations seen in the CMB spectra would clearly have an impact on the interpretation of a precision measurement of $B$ modes \cite{Ade:2014xna}.  
As can be seen from Fig.~\ref{fig:BB}, the gauge field can vary the height of the $BB$ spectrum at the reionization bump near $\ell \lesssim 10$ and at the primary acoustic peak near $\ell \sim 100$ by as much as $\pm 50\%$. However, we have the greatest leverage on new physics by focusing on the exotic cross correlations. Hence, we forecast the parameter constraints $\sigma_{R_{\rm YM}}$ and $\sigma_{g_{\rm YM}}$ using Fisher matrix techniques, for which the Fisher matrix reads
\begin{equation}
	\mathcal{F}_{ij}=\sum_l \sum_{X,Y}\frac{\partial C_l^X}{\partial \theta_i}\frac{\partial C_l^Y}{\partial \theta_j}\left[\Xi^{-1}_l\right]_{XY}
	\label{eqn:fisher}
\end{equation}
where $\vec \theta = (R_{\rm YM}, g_{\rm YM})$ and $X, \, Y = \{TB, EB\}$. The Fisher matrix $\mathcal{F}$ is the inverse of the covariance matrix between $R_{\rm YM}$ and $g_{\rm YM}$. The derivatives of the $C_l^X$ are obtained using CAMB. These derivatives are centered around a fiducial model: we choose $g_{\rm YM} = 10^{-56}$, $R_{\rm YM} = 0.1$ and the standard Planck $\Lambda$CDM values for the cosmological parameters \cite{Ade:2013zuv}. The matrix $\left[\Xi^{-1}_l\right]_{XY}$ is the inverse of the $TB$-$EB$ covariance matrix given by $\Xi ^{{X_1}{X_2},{X_3}{X_4}}_{l} = (\tilde C_l^{{X_1}{X_3}}\tilde C_l^{{X_2}{X_4}} + \tilde C_l^{{X_1}{X_4}}\tilde C_l^{{X_2}{X_3}})/(2l+1)$ where $\tilde C_l^{XX'} \equiv C_l^{XX'} + w_{XX'}^{ - 1}|W_l^b{|^{ - 2}}$ and $X=\{T, \, E, \, B\}$ \cite{Gluscevic:2010vv}. The instrumental parameters enter the window function $W_l^b \simeq \exp \left(-l^2 \sigma^2_b /2\right)$ due to the beam width, and the instrumental noise $w_{XX}^{-1}$, where $w_{TT}^{ - 1} \equiv  4\pi \sigma _T^2/N_{\text{pix}}$ and $w_{EE}^{ - 1} = w_{BB}^{ - 1} \equiv  4\pi \sigma _P^2/N_{\text{pix}}$ with the cross correlation contributions vanishing as the noise in the polarization is assumed to have no correlation to the noise in the temperature. In the window function $\sigma_b \equiv \theta_{\rm FWHM}/\sqrt{8\ln 2}$ where the beam width is measured in radians. Similarly, the number of pixels is $N_{\rm pix} = 4\pi\theta_{\rm FWHM}^{-2}$ and $\sigma_T$ and $\sigma_P$ are the temperature and polarization pixel noise. These are given by $\sigma_T^2 = ({\rm NET})^2 N_{\rm pix}/t_{\rm obs}$ and $\sigma_P = \sqrt{2}\sigma_T$ with NET being the noise-equivalent temperature and $t_{\rm obs}$ being the observation time. The parameters are taken from Ref.~\cite{Gluscevic:2010vv, Planck:2013cta} and summarized in Tab.~\ref{tab:obs}.

\begin{table}[htbp]
    \begin{tabular}{l c c c}
    \hline
    Instrument \,\,& \,$\theta _{{\rm FWHM}}$ [arcmin] & \,NET \,$[\mu \text{K}\sqrt{\text{s}}]$\, &  \,$t_{\text{obs}}$ [y] \\ \hline
	 Planck & 7.1 & 45 & 2 \\
	 CV limited & 5 & 0 & 1.2 \\ 
    \hline
    \end{tabular}
\caption{Instrumental parameters for the two experiments considered in this paper.  The parameters are the beamwidth $\theta _{\rm{FWHM}}$, noise-equivalent temperature NET, and observation time $t_{\text{obs}}$.}
\label{tab:obs}
\end{table}

\begin{figure}[htbp]
\vspace{-0.5cm}
\includegraphics[width=1\linewidth]{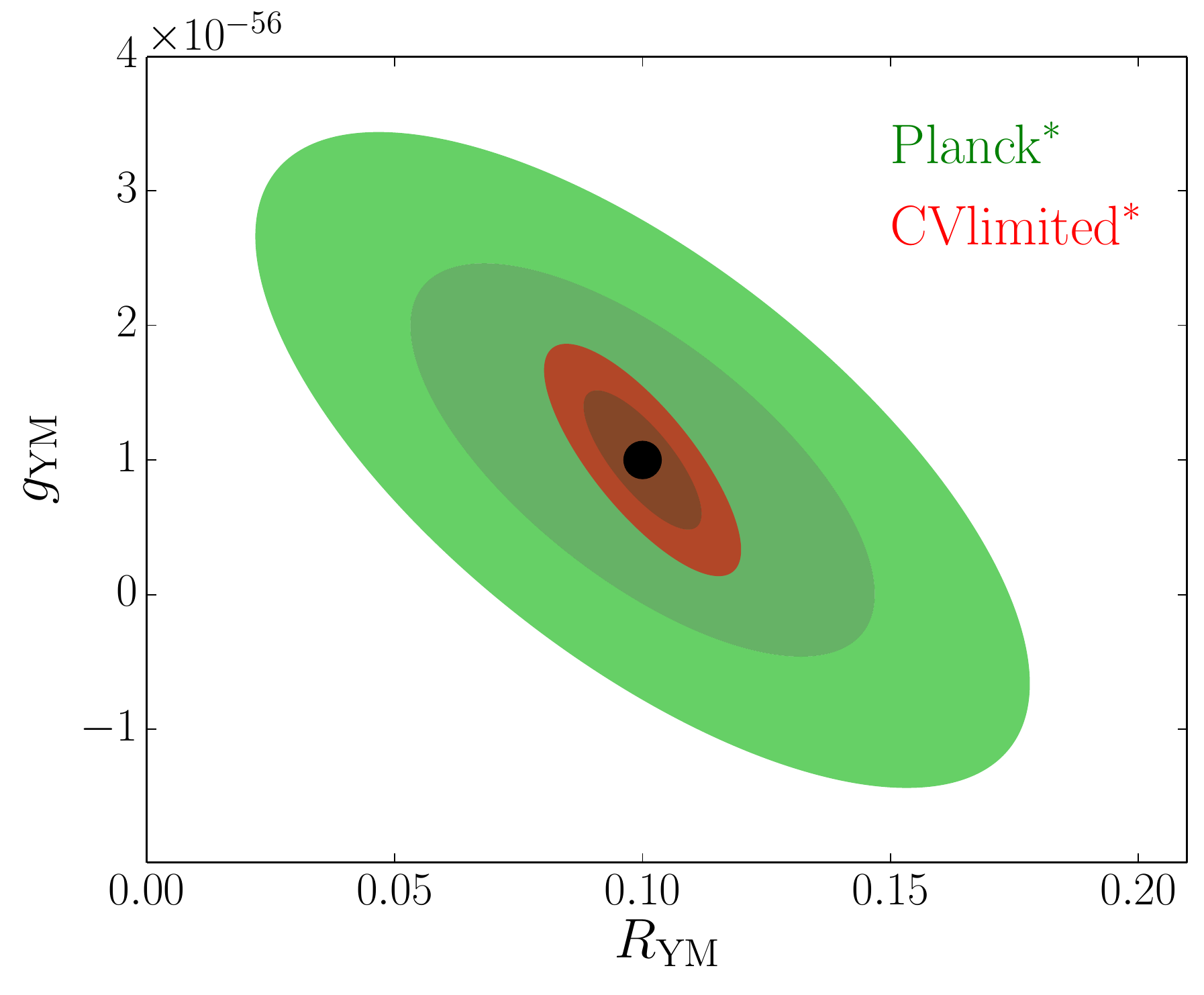}
\caption{Forecasted 1- and 2-$\sigma$ C.L. contours under the condition that $TB$ and $EB$ cross correlations are detected by the respective experiments (indicated by the asterisk $*$). The fiducial model is indicated by the black dot which represents $g_{\rm YM} = 10^{-56}$ and $R_{\rm YM} = 0.1$}
\label{fig:fisher}
\vspace{-0.25cm}
\end{figure}

The 1D marginalized confidence limits in a scenario in which the Planck satellite measures $TB$ and $EB$ correlations are $\sigma_{\rm g_{\rm YM}} = 9.5\times 10^{-57}$ and $\sigma_{R_{\rm YM}} = 0.030$. For the cosmic variance (CV) limited experiment these numbers reduce to $\sigma_{\rm g_{\rm YM}} = 3.4\times 10^{-57}$ and $\sigma_{R_{\rm YM}} = 8.1\times 10^{-3}$, which would be able to make a detection of $g_{\rm YM}$ feasible. The 1- and 2-$\sigma$ contours are plotted in Fig.~\ref{fig:fisher}. For this fiducial model Planck could make a 2-$\sigma$ detection of $R_{\rm YM}$, but cannot exclude the ambidextrous case since $g_{\rm YM} = 0$ lies within its 1-$\sigma$ contour. The future looks brighter for a future satellite mission that gets closer to a CV limited experiment, which could put constraints on the chiral asymmetry; for the fiducial model, the coupling $g_{\rm YM}$ could be distinguished from zero at better than the two sigma level. Such a limit could be used to determine whether the gauge field is part of a dark sector that includes dark energy. If dark energy couples to the rolling gauge field, 
or if the gauge field is dark energy, as in a gauge-flation scenario, then the rate of cosmic acceleration may be linked to the chiral asymmetry. 
 
Concluding, we present a simple model that breaks parity on cosmological scales. We illustrate the impact of this model on the gravitational wave spectrum and the CMB, computing the power spectra along with new $TB$ and $EB$ correlations that emerge in parity-breaking models. A detection of one of these correlations could be the sign of a flavor-space locked gauge field.

\acknowledgments
This work is supported in part by DOE grant DE-SC0010386. We thank Vera Gluscevic and Steve Gubser for useful conversations.
 

\end{document}